# A phenomenological model for magnetoresistance in granular polycrystalline colossal magnetoresistive materials: the role of spin polarised tunnelling at the grain boundaries

P. Raychaudhuri[‡], T. K. Nath, A. K. Nigam, R. Pinto

*Tata Institute of Fundamental Research,*

*Homi Bhaba Road, Colaba, Mumbai 400005, India.*

***Abstract:*** It has been observed that in bulk and polycrystalline thin films of collossal magnetoresistive (CMR) materials the magnetoresistance follows a different behaviour compared to single crystals or single crystalline films below the ferromagnetic transition temperature $T_c$. In this paper we develop a phenomenological model to explain the magnetic field dependence of resistance in granular CMR materials taking into account the spin polarised tunnelling at the grain boundaries. The model has been fitted to two systems, namely, $La_{0.55}Ho_{0.15}Sr_{0.3}MnO_3$ and $La_{1.8}Y_{0.5}Ca_{0.7}Mn_2O_7$. From the fitted result we have separated out, in $La_{0.55}Ho_{0.15}Sr_{0.3}MnO_3$, the intrinsic contribution from the intergranular contribution to the magnetoresistance coming from spin polarised tunnelling at the grain boundaries. It is observed that the temperature dependence of the intrinsic contribution to the magnetoresistance in $La_{0.55}Ho_{0.15}Sr_{0.3}MnO_3$ follows the prediction of double exchange model for all values of field.

[‡]e-mail:pratap@tifrc3.tifr.res.in

# I. Introduction

Hole doped $RMnO_3$ ( R = rare-earth ) type perovskites have attracted considerable attention in recent times because of their unusual magneto-transport properties arising from spin charge coupling. Hole doping in these materials is achieved by partially substituting the rare-earth ion (R) by a bivalent cation (M) like Ca, Sr or Ba. It has been observed that the compounds of the type $R_{1-x}M_xMnO_3$ like $La_{0.7}Sr_{0.3}MnO_3$ or $La_{0.7}Ca_{0.3}MnO_3$ exhibit very large magnetoresistance (MR~$\Delta\rho/\rho_0$=($\rho$(H)-$\rho$(H=0))/$\rho$(H=0))[1,2,3,4] near the ferromagnetic transition temperature. The large MR arises due to on site Hund's rule coupling between neighbouring $Mn^{3+}/Mn^{4+}$ pairs via Zener double exchange mechanism [5]. According to this mechanism the hopping probability of an electron between two adjacent $Mn^{3+}/Mn^{4+}$ is proportional to $\cos(\theta/2)$ where $\theta$ is the angle between the two manganese spins. Thus an electron has maximum mobility when the manganese ions are parallel to each other. An applied magnetic field suppresses the spin disorder of the manganese, aligning the manganese parallel to the field. This results in increased mobility of the electrons which in turn results in the drop of electrical resistance. Thus, below the ferromagnetic transition temperature $T_c$ one expects the MR to be simply related to the reduction in spin fluctuation in an applied magnetic field. However one actually observes a wide variety of field dependence depending on the microstructure of the material [6,7,8,9]. For granular polycrystalline samples much below $T_c$ one typically observes a sharp decrease in resistance at low fields followed by a slower almost linear decrease at higher fields. Similar behaviour is also observed in polycrystalline films grown on different substrates [10]. On the other hand for single crystals and single crystalline films the MR at low temperatures is very small and almost linear with magnetic field [9,10]. By comparing the magnetoresistance behaviour of the polycrystalline CMR material $La_{0.7}Ca_{0.3}MnO_3$ with different grain size Mahesh et al [11] have shown that the magnetoresistance of materials with smaller grain size is higher at temperatures below $T_c$ whereas the magnetoresistance at $T_c$ does not vary significantly. These results suggest that scattering at the polycrystalline grain boundaries play a significant role in determining the magnetoresistance in these materials in the ferromagnetic regime. Understanding the mechanism of magnetoresistance in granular polycrystalline materials is important since these materials have larger potential application due to their large magnetoresistance at low fields.

By comparing resistance versus field (R-H) data on polycrystalline bulk and single crystal of the same material Hwang et al [9] suggested that the low field magnetoresistance in polycrystalline materials is governed by the spin polarised transport across grain boundaries. One reason why they argued that spin polarised transport should be significant in these compounds is the high degree of spin polarisation. In perovskite manganites the relatively narrow majority carrier conduction band ( ~1.5 eV ) is completely separated from the minority band by a large Hund's rule as well as exchange energy ( ~2.5 eV ) leading to a complete polarisation of the conduction electrons [12]. By comparing the magnetisation as a function of field (M-H) with the R-H data on single crystals Hwang et al further suggested that scattering at magnetic domain boundaries in a single crystal is insignificant. Thus in this picture the low field drop in the R-H curve comes due to progressive alignment of the magnetic domains associated with the grains by the movement of domain walls across the grain boundaries.

Extending this idea further we develop, in this paper, a model which describes the magnetic field dependence of MR taking into account the gradual slippage of domain walls across the grain boundaries pinning centres in an applied magnetic field. The model is described in section 2. In section 3 we fit the model to two CMR manganites i) $La_{0.55}Ho_{0.15}Sr_{0.3}MnO_3$ which has pseudo-perovskite structure and ii) $La_{1.8}Y_{0.5}Ca_{0.7}Mn_2O_7$ which has a highly anisotropic layered perovskite structure. An attempt is made to separate out the intrinsic contribution to the magnetoresistance in $La_{0.55}Ho_{0.15}Sr_{0.3}MnO_3$, from the contribution coming from intergranular spin polarised transport. We have pointed out the need to separate out the intrinsic contribution from the intergranular contribution in any transport measurement on bulk samples before attempting to fit the data with any theoretical model based on the double exchange mechanism.

## II. Description of the model

Ferromagnets have an easy axis depending on the local crystallographic symmetry along which it is energetically favourable for the ferromagnetic spins to get aligned. Unlike the spins inside the domains which tend to get aligned along the easy axis, in a normal ferromagnet, the spins at the domain walls are at an angle with the easy axis which increases their anisotropy energy. Thus it is favourable for the domain wall to form at a defect site where the anisotropy energy is minimum due to the breaking of the local symmetry of the crystal. In polycrystalline materials the grain boundaries provide such pinning centres since the two adjacent grains have different anisotropy axes. Hence, in the absence of an applied magnetic field the domain wall tends to be in the grain boundary where it is pinned in a potential well, when the sample is cooled below the ferromagnetic transition temperature $T_c$. The free energy profile of the domain wall across grains and grain boundaries is schematically shown in figure 1a. Under an applied magnetic field the magnetisation reversal occurs through successive nucleation and propagation of the domain wall from the grain boundaries, with the field required to nucleate the domain boundaries higher than that required to propagate them. It is this mechanism of magnetisation reversal is known to give rise to Barkhausen jumps in many hard ferromagnets [13].

In polycrystalline CMR materials the spin polarised transport across a grain boundary will give a larger resistance when the two grains have misaligned magnetisation. In an applied magnetic field the domain wall experiences a force **f** ∝ **M$_S$·H,** where **M$_S$** is the spontaneous magnetisation and **H** is the applied magnetic field. When the magnetic field is large enough for domain boundary to overcome a grain boundary pinning well, that is, **f** ≥ [∇F(**r**)]$_{max}$ where F(**r**) is the free energy of the domain wall (figure 1b), the domain boundary moves out of the grain boundary giving rise to a drop in the electrical resistance of the material. In the present model we further simplify things by assuming the motion of domain walls to be in one dimension. The present model starts with the following assumptions:

(i) In zero field the domain boundaries are pinned at the grain boundary pinning centres.

The grain boundaries have a distribution of pinning strengths, $k \sim (1/M_s)dF(x)/dx$,

(defined as the minimum field needed to overcome a particular pinning barrier) given by $f(k)$.

(ii) When $H \geq k$ the domain boundary slips from the grain boundary giving rise to a resistance drop $\Delta r$. Thus the total drop in resistance due to spin polarised tunnelling at a field H is given by

$$\Delta R = N \Delta r \int_0^H f(k)\,dk \qquad -(1)$$

where N is the number of grain boundary domain walls initially present in the sample.

(iii) Following the observations of Hwang et al on single crystals we also assume that scattering at a magnetic domain boundary inside a grain is insignificant.

We further assume that the resistance has three parts; a magnetic field independent part $R_0$ coming from nonmagnetic defects and phonon scattering, a field dependent part coming from spin polarised tunnelling $R_{spt}(H)$ and a field dependent part coming from the reduction of spin fluctuation $R_{int}(H)$. Thus the total resistance R(H) can be written as

$$R(H) = R_0 + R_{spt}(H) + R_{int}(H).$$

The field dependence of the part coming from spin polarised tunnelling is given by (using (1))

$$R_{spt}(H) = R_{spt}(H=0)[1 - \int_0^H f(k)dk] \qquad -(2)$$

where $R_{spt}(H=0) = N\Delta r$. For the field dependence of $R_{int}(H)$ we rely on the experimental data on single crystals. It has been observed that the R-H curve is predominantly linear [9] with a weak higher order term appearing as one approaches $T_c$. We assume the field dependence to be

$$R_{int}(H) = -aH - bH^3 \qquad -(3)$$

with the second term being significant near the ferromagnetic transition temperature $T_c$. Using these we get the expression for magnetoresistance as,

$$MR = (R(H) - R(H=0))/R(H=0)$$

$$= -[R_{spt}(H=0)\int_0^H f(k)dk + aH + bH^3]/[R_0 + R_{spt}(H=0)]$$

$$= -A'\int_0^H f(k)dk - JH - KH^3, \qquad -(4)$$

where,

$A' = R_{spt}(H=0)/[R_0 + R_{spt}(H=0)]$,

$J = a/[R_0 + R_{spt}(H=0)]$,

and $K = b/[R_0 + R_{spt}(H=0)]$

are the fitting parameters.

Regarding the issue of which form of $f(k)$ would be most suitable is actually beyond the scope of the present study. However, we note that the sharp drop in the R-H curve is most pronounced at low values of field. It is thus reasonable to assume that there are grain boundary pinning centres of very weak strength $k$. We take $f(k)$ as a weighted average of a Gaussian and Skewed Gaussian distribution:

$$f(k) = A\exp(-Bk^2) + Ck^2\exp(-Dk^2) \quad -(5).$$

The fitting parameters finally are therefore A, B, C, D, J and K, with A′ absorbed in A and C.

Using this model we can now separate out the spin polarised intergranular contribution from the intrinsic contribution once the fitting parameters are found using equations (2) and (3). It might be noted here that as a first approximation Hwang et al [9] had earlier tried to estimate the contribution coming from spin polarised tunnelling by back-extrapolating the high field linear region of the MR-H curve to find the zero intercept. This method however fails at temperatures close to $T_c$ where the high field region no longer remains linear.

## III. Experimental Details

Magnetoresistance measurements were carried out on two CMR manganites $La_{0.55}Ho_{0.15}Sr_{0.3}MnO_3$ which has a $ABO_3$ type distorted perovskite structure and $La_{1.8}Y_{0.5}Ca_{0.7}Mn_2O_7$ ($T_c$~160 K) which is an electron doped [14] layered perovskite with two dimensional network of Mn-O-Mn bonds having a metal insulator transition temperature around 135 K. The polycrystalline samples were prepared through solid state reaction route starting from oxides of lanthanum, holmium, yttrium, and manganese and carbonates of strontium and calcium. Details of sample preparation are reported elsewhere [14,15]. We have earlier shown that with a small amount of holmium doping in $La_{0.7}Sr_{0.3}MnO_3$ the $T_c$ comes down [15] to suit the attainable temperature ranges of conventional low temperature cryostats. Thus for $La_{0.55}Ho_{0.15}Sr_{0.3}MnO_3$ ($T_c$~240K, *inset* figure 2) we measured the R-H curves from 5 K up to 230K. Magnetoresistance was measured using the conventional 4-probe technique in magnetic fields generated by a home made superconducting magnet.

## IV. Results and discussion

Figure 2 shows the MR as a function of field (MR-H) for various temperature as well as the fitted curves for the sample $La_{0.55}Ho_{0.15}Sr_{0.3}MnO_3$ . To fit equation (4) to these curves we used the following scheme. Differentiating equation (4) with respect to H and putting the expression (5) we get,

$$d(MR)/dH = A\exp(-BH^2) + CH^2\exp(-DH^2) - J - 3KH^2 \quad -(6).$$

The experimental curves in figure 2 were differentiated via cubic spline interpolation technique and fitted to equation (6) to find the best fit parameters. The inset of figure 3 shows the differentiated curve and the best fit function at 5K. Figure 3 shows the experimental MR-H curve along with the simulated one using equation (4). The excellent fit of the experimental data with the simulated curve shows that this procedure is self consistent.

Figure 2 shows the fitted curves at other temperatures using equation (4). There is an excellent fit for all temperatures up to 230 K where the spin polarised tunnelling contribution becomes zero. The coefficient of the cubic term K is significant only at 195 K. Using the expressions for $MR_{spt}(H)$ and $MR_{int}(H)$ we calculate the magnetoresistance coming from intergrain spin polarised tunnelling and the intrinsic contribution to the magnetoresistance at various temperature. Figure 4 shows the temperature variation of the total magnetoresistance, $MR_{int}(H)$ and $MR_{spt}(H)$ at 14 kOe respectively. We observe that the total magnetoresistance is a non-monotonic function of temperature with a slow decrease at low temperature followed by an increase as we approach $T_c$. The intrinsic contribution $MR_{int}(H=14kOe)$ however follows the expected double exchange behaviour with a steady increase with temperature. On the other hand $MR_{spt}(H=14 kOe)$ decreases steadily with temperature and totally vanishes at 230 K where the MR-H curve can be fitted with a cubic polynomial only. Hwang et al [9] had earlier observed that the temperature dependence of $MR_{spt}$ is described quite well by an expression of the type a+b/(c+T) which is a characteristic of spin polarised tunnelling in granular ferromagnetic systems. The inset of figure 4 shows the best fit of $MR_{spt}(H=14 kOe)$ with the expression a+b/(c+T). The fitted curve matches well with the extracted values of $MR_{spt}$ from the model. However our values of b and c for the best fit are much higher compared to that observed by Hwang et al though the $T_c$ of our system is much smaller. In this context we might note that the intergranular spin polarised tunnelling have different temperature dependence for ferromagnetically coupled and superparamagnetically coupled grains [16].

Figure 5 shows the MR-H curves along with the fitted curves (with equation (4)) for $La_{1.8}Y_{0.5}Ca_{0.7}Mn_2O_7$. In this case however we observe the appearance of a quadratic and cubic term in $MR_{int}(H)$ at relatively low temperatures (≥30 K). This might be related to the inherent two dimensionality of the magnetic lattice and is beyond the scope of the current paper.

## V. Conclusion

We have proposed a possible model for separating out the magnetoresistance arising from spin polarised transport from the intrinsic contribution in granular CMR materials. The model fits well with the experimental data on two systems, namely $La_{0.55}Ho_{0.15}Sr_{0.3}MnO_3$ and $La_{1.8}Y_{0.5}Ca_{0.7}Mn_2O_7$. The intrinsic contribution follows the behaviour expected from Zener double exchange mechanism. Since the polycrystalline grain boundaries are the primary source of spin polarised tunnelling studies of the low field behaviour for samples with controlled grain sizes will be highly interesting. Such work is currently under progress and will be published elsewhere.

## Acknowledgement

We would like to thank S. B. Roy, C. Mitra and S. Ramakrishnan for their keen interest in this work. We would also like to thank S. B. Roy, P. Chaddah, and S. Chaudhuri of Centre for Advanced Technology, Indore for their help regarding the magnetisation measurement on the SQUID magnetometer.

**References:**

1. R. von Helmolt, J. Weckerg, B. Holzapfel, L. Schultz, and K. Samwer, Phys. Rev. Lett., **71,** 2331 (1993)

2. R. Mahesh, R. Mahendiran, A. K. Raychaudhuri, and C. N. R. Rao, J. Solid State Chem., **114,** 297 (1995)

3. H. L. Ju, C. Kwon, Q. Li, R. L. Greene, and T. Venkatesan, Appl. Phys. Lett., **65,** 2109 (1994)

4. G. H. Jonker and J. H. Van Santen, Physica, **16,** 337 (1950)

5. C. Zener, Phys. Rev., **82,** 403 (1951)

6. H. L. Ju, J Gopalakrishnan, J. L. Peng, Q. Li, G. C. Xiong, T. Venkatesan, and R. L. Greene , Phys. Rev., **B51,** 6143 (1995)

7. P. Schiffer, A. P. Ramirez, W. Bao, and S-W Cheong, Phys. Rev. Lett., **75,** 3336 (1995) and see also A. P. Ramirez, J. Phys.: Condens. Matter, **9,** 8171 (1997)

8. A. Gupta, G. Q. Gong, G Xiao, P. R. Duncombe, P. Lecoeur, P. Trouilloud, Y. Y. Wang, V. P. Dravdi, and J. Z. Sun, Phys. Rev., **B54,** R15629 (1996)

9. H. Y. Hwang, S-W. Cheong, N. P. Ong, B. Batlogg, Phys. Rev. Lett., **77,** 2041 (1996)

10. R. Shreekala, M. Rajeswari, K. Ghosh, A. Goyal, J. Y. Gu, C. Kwon, Z. Trajanovic, T. Boettcher, R. L. Greene, R. Ramesh, and T. Venkatesan, Appl. Phys. Lett., **71,** 282 (1997)

11. R. Mahesh, R. Mahendiran, A. K. Raychaudhuri, C. N. R. Rao, Appl. Phys. Lett., **68,** 2291 (1996)

12. Park J-H, Chen C T, Cheong S-W, Bao W, Meigs G, Chakarian V, and Idzeral Y U, Phys. Rev. Lett., **76,** 4215 (1996)

13. P. J. Thompson and R. Street, J. Magn. Magn. Mater., **171,** 153 (1997)

14. P. Raychaudhuri, C. Mitra, A. Paramekanti, R. Pinto, A. K. Nigam, and S. K. Dhar, (to be published)

15. P. Raychaudhuri, T. K. Nath, P. Sinha, C. Mitra, A. K. Nigam, S. K. Dhar, and R. Pinto, J. Phys.: Condens. Mater., **9,** 10919 (1997)

16. J. S. Helman and B. Abeles, Phys. Rev. Lett., **37,** 1429 (1976)

# Figure Captions

**Figure 1:** (a) Free energy profile of a domain wall at grains and grain boundaries. (b) Expanded view showing the pinning strength as the maximum slope of the pinning well.

**Figure 2:** The experimental MR-H curves (dots) for $La_{0.55}Ho_{0.15}Sr_{0.3}MnO_3$ and the fitted curves (lines) using equation (4) at various temperatures; the *inset* shows the magnetisation as a function of temperature at 5000 kOe.

**Figure 3:** The magnetoresistance versus field (MR-H) curve at 5 K: the dots (•) are the experimental points and the line (−) is the fitted curve using equation (4). The *inset* shows the derivative of the experimental curve (•) and the fitted curve (−) using the equation $d(MR)/dH = A\exp(-BH^2) + CH^2\exp(-DH^2) - J - 3KH^2$, with A=-0.2678, B=2.1665, C=-0.0125, D=0.4300, J=2.73×10$^{-3}$ (see text).

**Figure 4:** Temperature dependence of various components of magnetoresistance (MR) at 14 kOe : (♦) is the total MR; (•) is the spin polarised contribution to the magnetoresistance ($MR_{spt}$); (▲) is the intrinsic contribution $MR_{int}$. The *inset* shows the best fit of $MR_{spt}$ to a function of the form $a+b/(c+T)$, with a=-0.3468, b=238.5 K, c=446.5 K (see text).

**Figure 5:** The experimental MR-H curves (dots) for $La_{1.8}Y_{0.5}Ca_{0.7}Mn_2O_7$ along with the fitted curve using equation (4) (broken lines) at 5 K, 60 K, 85 K and 100 K; The spin polarised tunnelling contribution at 14 kOe ( $MR_{spt}$(H=14 kOe) ) at these four temperatures are 0.2634, 0.2162, 0.1583 and 0.1419 respectively.

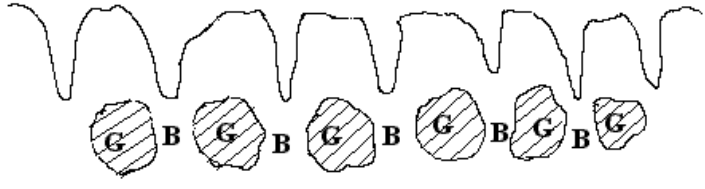

(a)

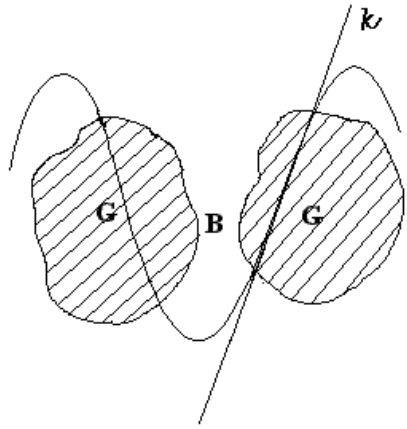

(b)

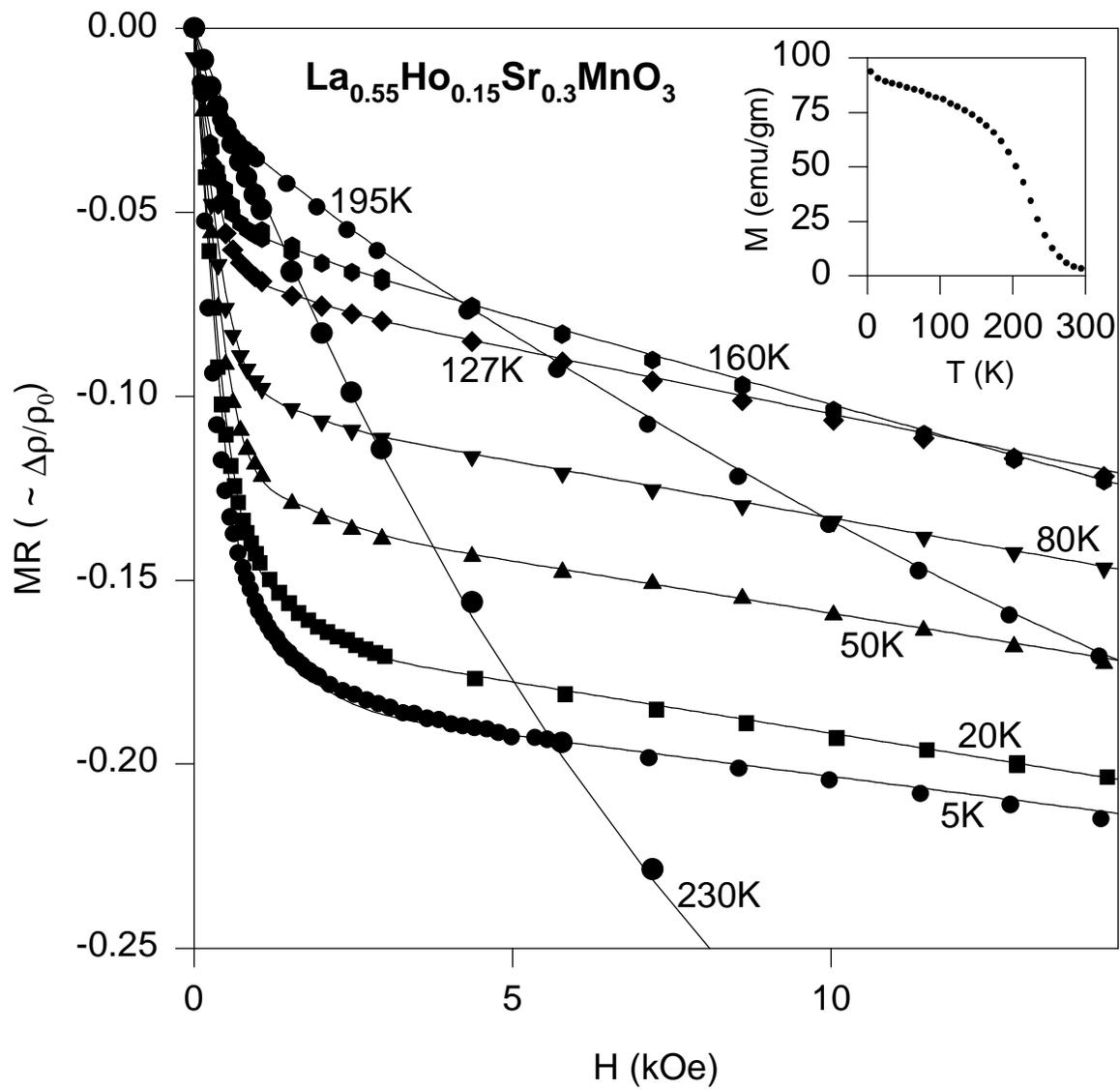

Figure 2 (P Raychaudhuri et al)

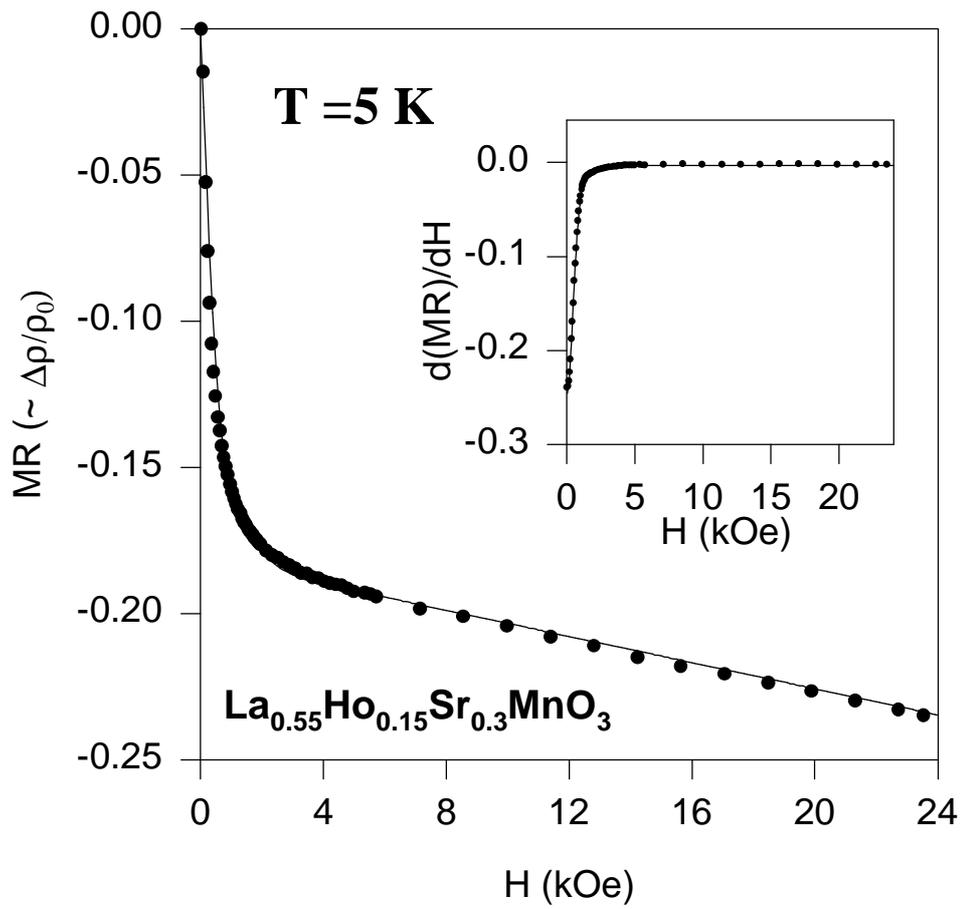

Figure 3 (P Raychaudhuri et al)

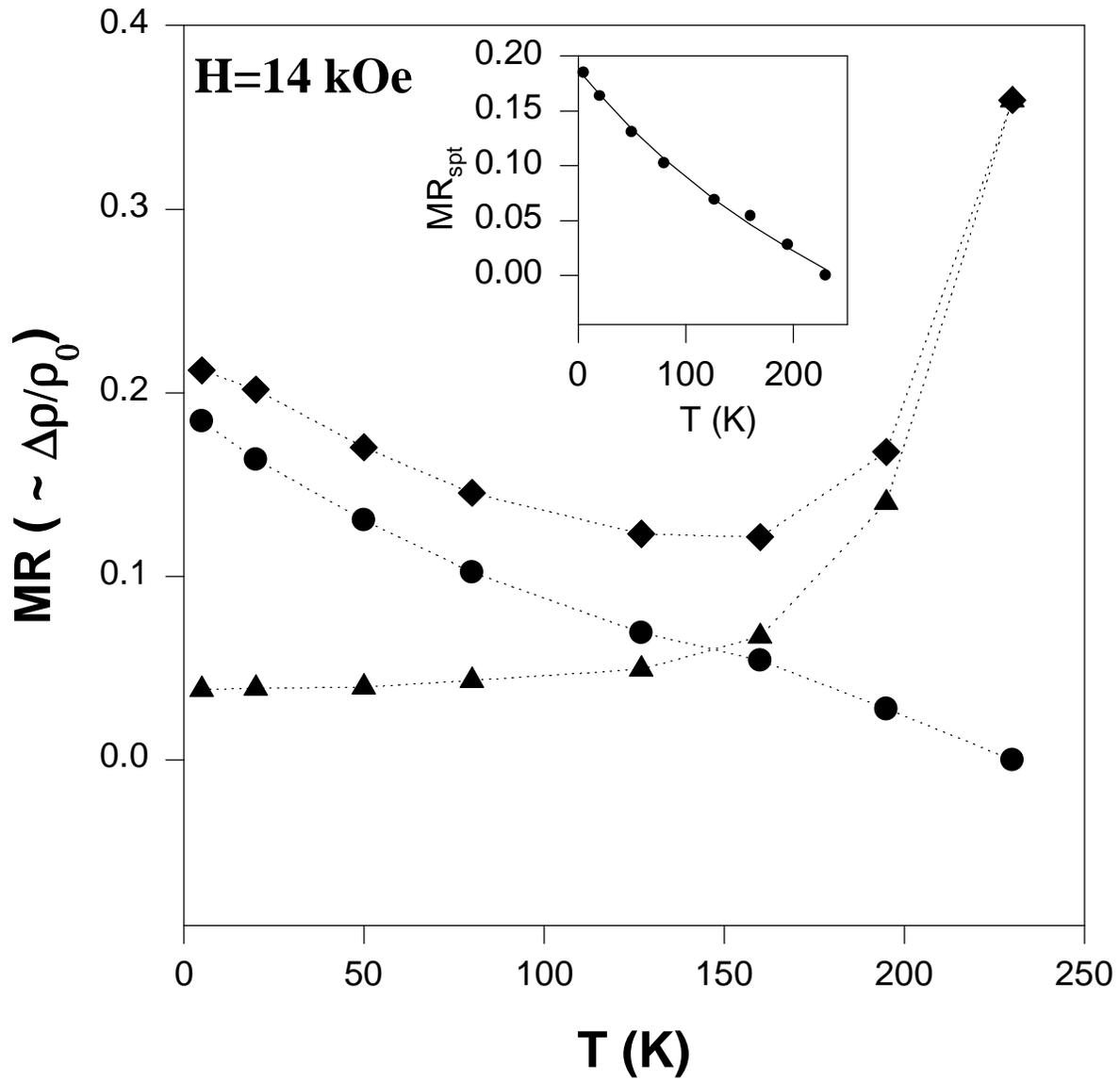

Figure 4 (P Raychaudhuri)

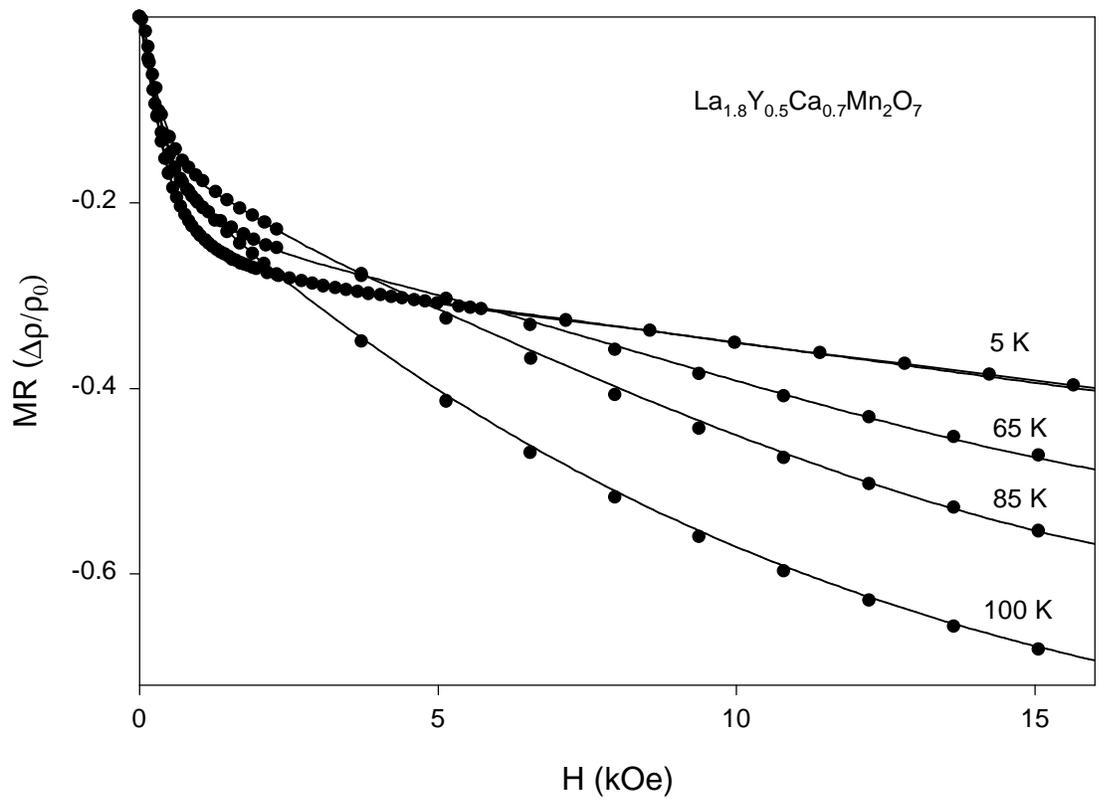